\documentclass[a4paper]{PoS}
\usepackage[numbers]{natbib}
\usepackage{psfig}
\usepackage{graphicx}

\title{Exploring Multiwavelength AGN Variability with Swift Archival Data}

\ShortTitle{Exploring Multiwavelength AGN Variability with Swift Archival Data}

\author{\speaker{Jonathan Gelbord},$^a$ Caryl Gronwall,$^b$ Dirk
  Grupe,$^c$ Dan Vanden Berk$^d$ and Jian Wu$^e$\\
  \llap{$^a$}Spectral Sciences Inc., 4 Fourth Ave., Burlington MA
  01803, USA\\
  \llap{$^b$}Department of Astronomy and Astrophysics, Pennsylvania
  State University, 525 Davey Lab, University Park PA 16802, USA\\
  \llap{$^c$}Department of Earth and Space Sciences, Morehead State
  University, 235 Martindale Drive, Morehead KY 40351, USA\\
  \llap{$^d$}Department of Physics, St.\ Vincent College, 300 Fraser
  Purchase Rd, Latrobe PA 15650, USA\\
  \llap{$^e$}College of Information Sciences and Technology,
  Pennsylvania State University, 322 IST, University Park PA 16802, USA\\
  E-mail: \email{jgelbord@spectral.com}, \email{jxw394@ist.psu.edu}}

\abstract{We are conducting an archival {\it Swift} program to measure
  multiwavelength variability in active galactic nuclei (AGN).  This
  variability information will provide constraints on the geometry,
  physical conditions and processes of the structures around the
  central black holes that emit and reprocess the observed flux.
  Among our goals are: 
  (1) to produce a catalog of type 1 AGN with time-resolved
  multi-wavelength data; 
  (2) to characterize variability in the optical, UV and X-ay bands as
  well as changes in spectral slope; 
  (3) to quantify the impact of variability on multi-wavelength
  properties;
  and
  (4) to measure correlated variability between bands.
  Our initial efforts have revealed a UVOT calibration issue that can
  cause a few percent of measured UV fluxes to be anomalously low, by
  up to 30\%.}

\FullConference{Swift: 10 Years of Discovery,\\
		2-5 December 2014\\
		La Sapienza University, Rome, Italy }

\begin{document}

\section{Introduction}
\label{sec:intro}

\vspace*{-2mm}
\subsection{Multiwavelength variability in AGN}
\label{sec:introVar}
Active Galactic Nuclei (AGN) are intrinsically broad-band beasts, with
emission spanning the electromagnetic spectrum \citep[e.g.,][]{elvis94}.
Emission in different wavebands is in general dominated by different
structures.  For instance, the optical-to-ultraviolet (UV) emission is
dominated by the big blue bump (BBB), which is the integrated black
body emission from regions of the accretion disk with a distribution
of temperatures, whilst X-rays are thought to be due to inverse
Compton (IC) scattering (likely the upscattering of BBB photons by a
coronal region around the accretion disk), and the infrared (IR) band
exhibits thermal emission from both AGN-heated dust and the host
galaxy stellar population.
Consequently, a complete picture of AGN
requires multiwavelength data.  When parts of the spectral energy
distribution (SED) are not observed, a generic spectral shape is
assumed to estimate the unobserved flux.

\begin{figure}[bt]
  \parbox{\textwidth}
     {
       \centering
       \rotatebox{0}{\resizebox{0.45\textwidth}{!}
         {\includegraphics[width=0.45\textwidth]{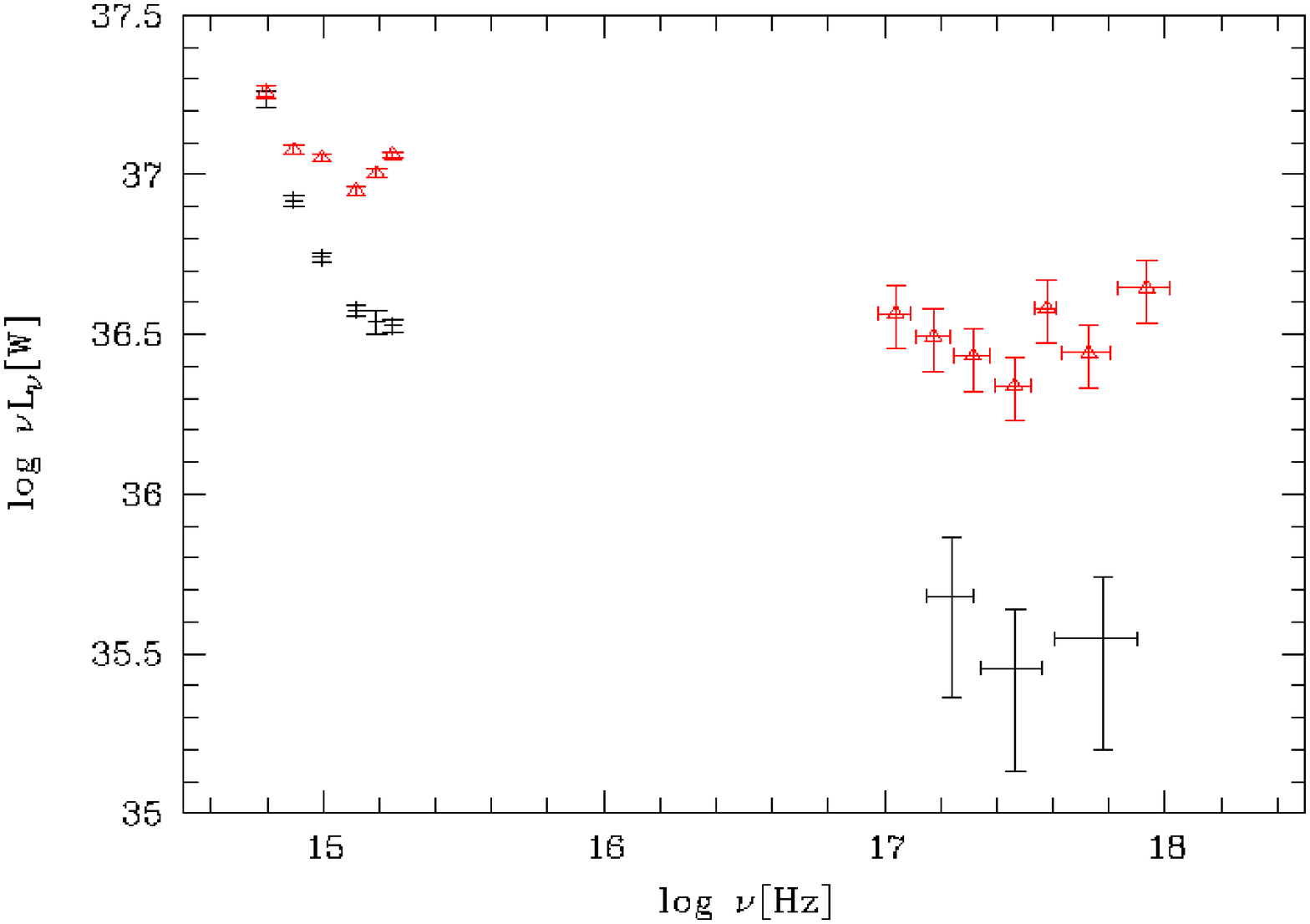}}}
     }
   \begin{minipage}[b]{\textwidth}
     \centering
     \caption{Optical through X-ray SEDs for 1RXS J2349-31 as measured
       by {\it Swift} in September 2006 (black) and May 2007 (red).
       The X-ray flux increased tenfold whereas the optical-UV fluxes
       rose by no more than a factor of three, causing
       $\alpha_\mathrm{ox}$ to change from $-1.50$ to $-1.25$ in eight
       months \citep{grupe10}.
       \vspace*{-2mm}}
     \label{fig:J2349}
   \end{minipage}
\end{figure}

AGN emission is also variable, with intrinsic time scales that may
differ for each emitting region.
The most compact regions, generally the energetic regions closest to
the central black hole, can vary the most rapidly.
Consequently, the temporal behavior can differ across the SED.
If the timescale or amplitude of variations differ between wavebands, 
this will unavoidably influence multiwavelength measurements
such as the optical to X-ray spectral index, $\alpha_\mathrm{ox}$
(defined as $f_\nu \propto \nu^{+\alpha}$; effectively the relative
power of the accretion disk and the IC emission).  
As a result, a single measurement may not provide a
typical value for a given AGN (e.g., Figure~\ref{fig:J2349}, in which
the $\alpha_\mathrm{ox}$ of 1RXS J2349--31 is shown to change by
$+0.25$ within a span of eight months).
Variability therefore adds noise to measurements of multiwavelength
parameters, which will weaken or obscure correlations that could
provide insight into the properties and processes of AGN.

Historically, the availability of simultaneous multiwavelength data
has been limited due to the difficulty of coordinating observations.
As a result, non-simultaneous data have often been combined to
establish multiwavelength parameters.  The danger of this is
demonstrated by 1RXS J2349--31 (Figure~\ref{fig:J2349}): if the
optical measurements of one epoch are combined with the X-ray
measurements of the other, the $\alpha_\mathrm{ox}$ values obtained
would be either $-1.75$ or $-1.00$, whereas measurements from
simultaneous data only show a change from $-1.50$ to $-1.25$.
Instances such as this amplify the impact of variability upon
multiwavelength parameter measurements when non-simultaneous data are
combined.
For example,
an anti-correlation has been established between UV luminosity
(L$_\mathrm{UV}$) and $\alpha_\mathrm{ox}$ \citep[e.g.,][]{just07}.
We have shown that the scatter in this correlation is reduced by at
least 20\% when only simultaneous UV and X-ray data are used
\citep{wu12}.

\vspace*{-2mm}
\subsection{Why \emph{Swift}?}
\label{sec:introSwift}
The data archive of the \emph{Swift} satellite is a unique resource
for studies of multiwavelength variability.
By default, \emph{Swift} collects data simultaneously with the X-ray
Telescope (XRT) and the Ultraviolet/Optical Telescope (UVOT).  
The availability of simultaneous data in these bands is critical, as
these are the bands in which radio-quiet AGN vary most strongly.

The XRT and UVOT have wide, co-aligned fields of view, providing
nearly 300 sq.\ arcmin of overlapping area with each pointing.
This large area provides opportunities to measure many serendipitous
sources as well as the intended targets of each observation.
As of 13 May 2012, the \emph{Swift} archive included 667 sq.\ degrees
of sky covered by both instruments with exposure times of at least
1~ks and 222 sq.\ degrees with at least 10~ks of exposure time.
In addition, there are many fields with repeated observations, making
it possible to measure variability on time scales ranging from hours
to years.  This includes 400 sq.\ degrees with data taken at
least one day apart.

\emph{XMM-Newton} also offers simultaneous X-ray, UV and optical data.
In contrast to \emph{Swift}, which is in a 1.6 hour orbit and cannot
observe any point in the sky for more than $\sim$ 40 min per
orbit, \emph{XMM} is capable of observing a single field continuously
for more than a day, so it is better suited for measuring 
variability behavior on intraday timescales.
However, for timescales of days to years, \emph{Swift} provides
unparalleled temporal coverage, with vastly more repeated fields and
higher cadence monitoring (Table~\ref{tab:repeats}).

\begin{table}[bt]
\centering
\begin{tabular}{c@{\ \ \ \ \ \ }c@{\ \ \ \ \ \ }c}
Min.\ num.\ of days with data  &  \emph{Swift} fields  &  \emph{XMM} fields  \\ \hline
  2                   &  5000                 &  736  \\
  20                  &  350                  &  6    \\
  50                  &  150                  &  0    \\
  100                 &  44                   &  0    \\
\end{tabular}
\caption{Tallies of fields that were re-observed on multiple days by
  \emph{Swift} and \emph{XMM}.  For both observatories, the maximum
  area per field with simultaneous X-ray and UV or optical data is
  $17' \times 17'$.  \emph{Swift} data are as of 13 May 2012 and
  \emph{XMM} data are based on the 2XMM-DR3 catalog \citep{watson09}.
  \vspace*{-2mm}} 
\label{tab:repeats}
\end{table}

\section{The catalog}

We are currently building a catalog of AGN with multiple multiwavelength
observations in the \emph{Swift} archive.  The sample will be limited
to spectroscopically-confirmed AGN with secure classifications,
redshifts and means of estimating parameters such as their black hole
masses (M$_\mathrm{BH}$) and Eddington fractions (L/L$_\mathrm{Edd}$).
We will compile X-ray data and photometry from six UVOT filters.
The catalog will include time-resolved flux values, upper limits for
epochs with non-detections, spectral indices and hardness ratios, and
estimates of the fluxes in fixed bands in the emitted frame.

We anticipate having 2000--4000 AGN in the final sample.  This will
include about 2000 SDSS QSOs from the SDSS-DR7, -DR9 and -DR10 quasar
catalogs \citep{schneider10, paris12, paris14} and up to 2000
additional AGN appearing in the literature.  Of order 1000 of the
sample members will be purely random, serendipitously-observed
sources.  The sample
will represent a broad range of intrinsic properties (such as
luminosities, black hole masses and accretion rates) and will be large
enough to subdivide by these parameters in order to assess how they
affect the variability behavior.

\section{Objectives}

The primary goals of this program may be summarized as follows:
\begin{itemize}

\item To build the largest catalog of AGN with simultaneous X-ray,
  optical and UV data.

\item To use these data to obtain cleaner measurements of
  multiwavelength parameters, with which we will refine established
  correlations and test for new ones.

\item To quantify the flux and spectral variability on timescales from
  hours to years, characterizing both individual objects and average
  behavior amongst subsamples defined by various intrinsic properties.

\item For a limited number of AGN with both intensive monitoring and
  high signal-to-noise ratio data, we will put constraints on
  any correlated variability between the bands.

\end{itemize}

Taken together, these efforts will quantify the strength and rate of
incidence of spectral variability, produce improved estimates of AGN
bolometric luminosities, and provide constraints on the geometry and
properties of the emission region structures.  Any relationships found
between observed variability and intrinsic parameters should provide
insight into the underlying processes.

\section{Spurious UVOT light curve dips: a newly discovered calibration issue}

\begin{figure}[bt]
  \parbox{\textwidth}
     {
       \centering
       \rotatebox{0}{\resizebox{0.6\textwidth}{!}
         {\includegraphics*[119,174][497,626]{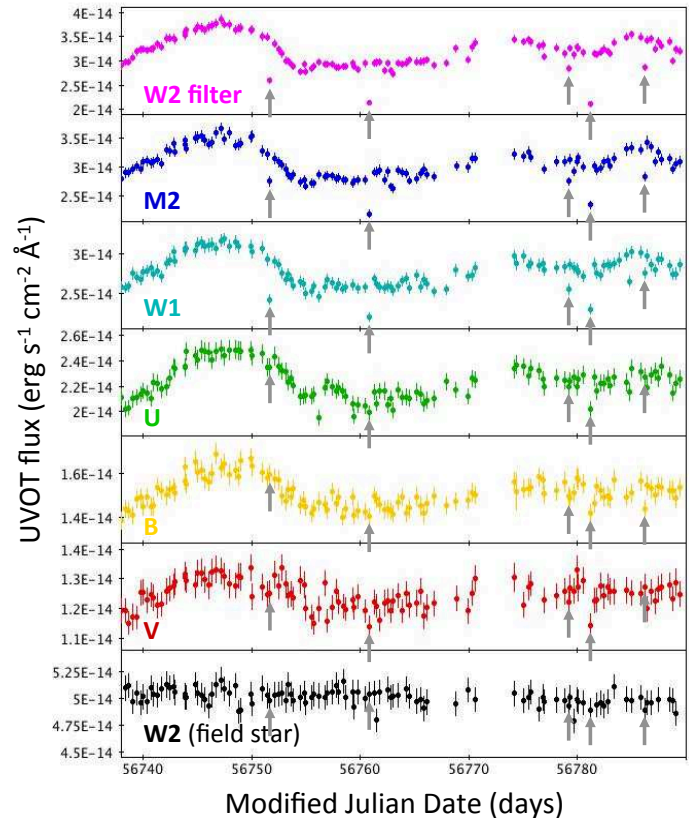}}}
     }
   \begin{minipage}[b]{\textwidth}
     \centering
     \caption{UVOT light curves for NGC 5548 in six filters
       (increasing in wavelength from the top panel down).  For
       comparison, the bottom panel presents the W2 light curve of a
       field star.  Grey arrows indicate measurements from five
       observations in which the AGN light curve exhibits clear dips
       in the UV filters.
       \vspace*{-2mm}}
     \label{fig:UVOT_LC}
   \end{minipage}
\end{figure}

As a test case for our UVOT data processing scripts, we produced light
curves for the AGN NGC 5548 \citep{mchardy14,edelson15}.
In Figure~\ref{fig:UVOT_LC}, we present a section of these light
curves during which each observation included data in all six
filters.
A striking feature of these curves is a set of abrupt,
short-lived flux dips.
These apparent flux drops are 
(1) seen in simultaneous data across multiple filters; 
(2) strongest at short wavelengths (by up to 34\% in UVW2, 22\% in UVM2
and 17\% in UVW1); 
(3) weak or absent in the optical bands; 
(4) found to exceed 10\% in 4.5\% of UVW2 measurements, 3.3\% of UVM2,
1.0\% of UVW1 and never in the optical filters; 
(5) sporadic, not periodic; and
(6) not seen in the light curves of any field stars.

Upon first inspection, the light curve dips look real.  
There are no obvious defects in the data (no image artifacts, elevated
backgrounds, tracking errors, known bad pixels, etc.).  
Moreover, no other objects in the field are affected when the AGN
dips.
However, the short timescales are physically implausible (extreme
examples include a 33\% drop in 29 min and a 20\% drop in just 19 min;
in both instances, fluxes measured hours later were consistent with
the levels before the drops).
Additionally, no comparable dips are found in contemporaneous UV
data from \emph{HST} \citep{derosa15}.

\begin{figure}[bt]
  \parbox{\textwidth}
     {
       \centering
       \rotatebox{0}{\resizebox{0.9\textwidth}{!}
         {\includegraphics*[41,318][569,475]{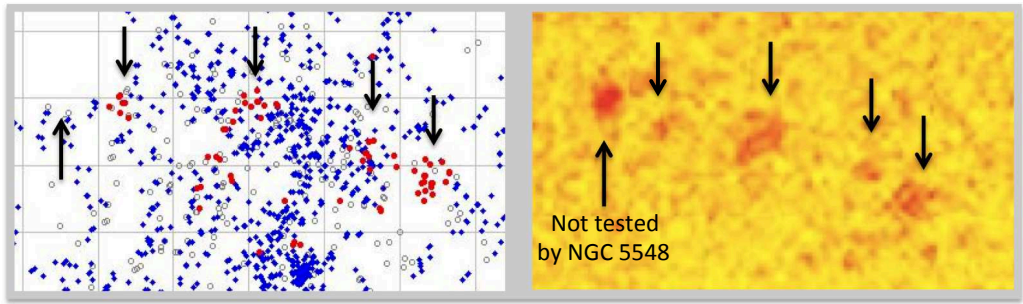}}}
     }
   \begin{minipage}[b]{\textwidth}
     \centering
     \caption{A comparison of the spatial distribution of NGC 5548
       flux outliers to blemishes present in a source-subtracted
       background image in the UVOT UVM2 filter.  At left, we display
       a map of NGC 5548 measurements in raw UVOT detector coordinates
       (spanning roughly from X,Y = 295,505 to 620,710 in a frame with
       1024 $\times$ 1024 pixels [i.e., the default 2$\times$ binning]).
       Red points are flux dips, blue points are non-dipping
       measurements and open circles are measurements for which the
       dipping test was not applied because 
       of insufficient data within $\pm$2 days.
       At right is a source-subtracted background image in the UVOT M2
       filter covering the same detector region \citep{breeveld10}.
       This image exhibits dark patches where the measured background
       flux is lower.  The locations of these blemishes are indicated
       by arrows in both panels.
       \vspace*{-2mm}}
     \label{fig:Blemishes}
   \end{minipage}
\end{figure}

By comparing measured fluxes to neighboring points in the light curve,
we have identified 85 dips \citep{edelson15}.  Projecting these back
into raw detector coordinates, we found 
that almost all are tightly
clustered within a few small regions on the detector
(Figure~\ref{fig:Blemishes}, left panel).  
It appears that these are detector locations at which the
sensitivity is reduced, especially at short wavelengths.
While this effect has not been noted previously, these regions line up
with features that can be seen in the source-subtracted background
images presented by Breeveld et al.\ (\citep{breeveld10}; 
Figure~\ref{fig:Blemishes}, right panel).

In our analysis of the dips in the NGC 5548 light curve, we define
regions within which the detector sensitivity appears to be reduced
(see \citep{edelson15} for details).
These preliminary regions are adequate to clean up the NGC 5548 light
curves, but the data from this AGN alone are insufficient to precisely
define all areas of reduced sensitivity; portions of the defined
regions may not have reduced sensitivity whilst there may be other
problematic areas that were not probed by the NGC 5548 data.
However, the fact that the present regions cover 4\% of the central 5'
$\times$ 5' of the UVOT field of view suggests that a few percent of
all UVOT measurements may be affected.

\section*{Acknowledgements}
We gratefully acknowledge the support from NASA under award NNH13CH61C.


\begin{thebibliography}{99}

\bibitem[\protect\citeauthoryear{Breeveld et al.}{2010}]{breeveld10}
  Breeveld, A.~A., et al., 2010
  \emph{Further Calibration of the Swift Ultraviolet/Optical Telescope},
  \emph{MNRAS} {\bf 406}, 1687.

\bibitem[\protect\citeauthoryear{De Rosa et al.}{2015}]{derosa15}
  De Rosa, G., et al., 2015
  \emph{Space Telescope and Optical Reverberation Mapping
    Project. I. Ultraviolet Observations of the Seyfert 1 Galaxy NGC
    5548 with the Cosmic Origins Spectrograph on Hubble Space Telescope},
  \emph{accepted to ApJ} {\tt (arXiv:1501.05954)}.

\bibitem[\protect\citeauthoryear{Edelson et al.}{2015}]{edelson15}
  Edelson, R., et al., 2015
  \emph{Space Telescope and Optical Reverberation Mapping
    Project. II. Swift and HST Reverberation Mapping of the Accretion
    Disk of NGC 5548},
  \emph{accepted to ApJ} {\tt (arXiv:1501.05951)}.

\bibitem[\protect\citeauthoryear{Elvis et al.}{1994}]{elvis94}
  Elvis, M., et al., 1994
  \emph{Atlas of Quasar Energy Distributions},
  \emph{ApJS} {\bf 95}, 1.

\bibitem[\protect\citeauthoryear{Grupe et al.}{2010}]{grupe10}
  Grupe, D., et al., 2010
  \emph{The Simultaneous Optical-to-X-ray Spectral Energy Distribution
  of Soft X-ray Selected Active Galactic Nuclei Observed by Swift},
  \emph{ApJS} {\bf 187}, 64.

\bibitem[\protect\citeauthoryear{Just et al.}{2007}]{just07}
  Just, D.W., et al., 2007
  \emph{The X-Ray Properties of the Most Luminous Quasars from the
    Sloan Digital Sky Survey},
  \emph{ApJ} {\bf 665}, 1004.

\bibitem[\protect\citeauthoryear{McHardy et al.}{2014}]{mchardy14}
  McHardy, I.~M., et al., 2014
  \emph{Swift Monitoring of NGC 5548: X-ray Reprocessing and
    Short-term UV/Optical Variability},
  \emph{MNRAS} {\bf 444}, 1469.

\bibitem[\protect\citeauthoryear{P{\^a}ris et al.}{2012}]{paris12}
  P{\^a}ris, I., et al., 2012
  \emph{The SDSS Quasar Catalog: Ninth Data Release},
  \emph{A\&A} {\bf 548}, 66.

\bibitem[\protect\citeauthoryear{P{\^a}ris et al.}{2014}]{paris14}
  P{\^a}ris, I., et al., 2014
  \emph{The SDSS Quasar Catalog: Tenth Data Release},
  \emph{A\&A} {\bf 563}, 54.

\bibitem[\protect\citeauthoryear{Schneider et al.}{2010}]{schneider10}
  Schneider, D.P., et al., 2010
  \emph{The SDSS Quasar Catalog. V. Seventh Data Release},
  \emph{AJ} {\bf 139}, 2360.

\bibitem[\protect\citeauthoryear{Watson et al.}{2009}]{watson09}
  Watson, M., et al., 2009
  \emph{The XMM-Newton serendipitous survey. V. The Second XMM-Newton
    serendipitous source catalogue},
  \emph{A\&A} {\bf 493}, 339.

\bibitem[\protect\citeauthoryear{Wu et al.}{2012}]{wu12}
  Wu, J., et al., 2012
  \emph{A Quasar Catalog with Simultaneous UV, Optical, and X-Ray Observations by Swift},
  \emph{ApJS} {\bf 201}, 10.

\end{thebibliography}
\end{document}